\documentclass[a4paper,11pt]{article}
\usepackage[dvipdfmx]{graphicx}
\usepackage{pos}

\title{Characterization of SiPM and development of test bench modules for the next-generation cameras for Large-Sized Telescopes for Cherenkov Telescope Array}
 \ShortTitle{Development of SiPM modules for CTA-LST}

\author*[a]{T. Saito}
\author[a]{K. Hashiyama}
\author[b]{H. Iwasaki}
\author[a]{H. Kubo}
\author[c]{M. Mizote}
\author[d]{A. Okumura}
\author[d]{H. Tajima}
\author[c]{T. Yamamoto} \onbehalf{on behalf of the CTA-LST Project}
\affiliation[a]{Institute for Cosmic Ray Research, the university of Tokyo, Kashiwanoha 5-1-5, Kashiwa, Japan}
\affiliation[b]{Kyoto University, Kitashirakawa Oiwakecho, Kyoto, Japan}
\affiliation[c]{Konan University, Okamoto 8-9-1, Higashinada-ku, Kobe, Japan}
\affiliation[d]{ISEE, Nagoya University, Chikusa-ku, Nagoya, Japan}

\emailAdd{tsaito@icrr.u-tokyo.ac.jp}
\abstract{The recent improvements in the performance of the silicon photomultipliers (SiPMs) made them attractive options as photo sensors of imaging atmospheric Cherenkov telescopes (IACTs). In fact, they are already adopted in some IACTs such as FACT and the Small-Sized Telescopes of the Cherenkov Telescope Array (CTA). However, the application to the Large-Sized Telescopes (LSTs) of CTA requires additional studies. As the pixel size of LSTs is larger than the nominal size of SiPMs, the signal from multiple sensors must be summed up. Also, the high detection efficiency of the night sky background (NSB) photons may degrade the telescope performance. To overcome this, the pulse width must be as small as 3 ns and the detection efficiency for NSB photons must be suppressed as much as possible. Heat generation and gain stabilization are also issues. We studied different types of SiPMs from Hamamatsu photonics and characterized them for the LST application, addressing the previous points. Also, to prove the SiPM performance in LST, we are developing a SiPM module which can be installed in the exisiting LST camera. Here we present the results of this evaluation and the status of the test bench module development.}

\ConferenceLogo{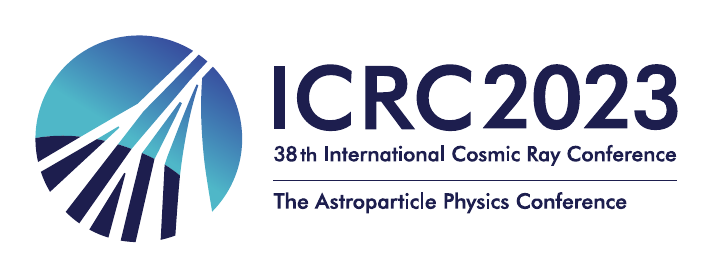}

\FullConference{%
38th International Cosmic Ray Conference (ICRC2023)\\
  26 July - 3 August, 2023\\
  Nagoya, Japan}

\begin{document}
\maketitle

\section{Large-Size Telescopes for Cherenkov Telescope Array}
The Cherenkov Telescope Array (CTA) is the next generation gamma-ray observatory, covering the energy range from 20 GeV to 300 TeV. Such a wide energy coverage can be achieved by using three different sizes of Imaging Atmospheric Cherenkov Telescopes (IACTs), namely, Large-Sized Telescope (LST, the reflector diameter $D_r =23$~m), Midium-Sized Telescope (MST, $D_r = \sim 10$~m)) , and Small-Sized Telescope (SST, $D_r = \sim 4$~m). 
The two arrays, one in each hemisphere to survey the entire sky, will include a combination of many SSTs, observing very high energy gamma-rays above 3 TeV, several MSTs taking charge of the middle energy range between 300 GeV to 10 TeV and LSTs, which are sensitive to the lowest energy range above 20 GeV. 

 Cherenkov light from an atmospheric air shower is dominated by optical photons and IACTs are designed to catch them. As photo sensors, conventional photomultipliers (PMTs) have been used in IACTs since the beginning of the field.  Semiconductor type photosensors, so-called Silicon Photomultiplers (SiPMs), have also started to be used recently. The FACT telescope \cite{FACT} was the first IACT, that was equipped with SiPMs. It started operation in 2012. The sensitivity of SiPM has been getting higher, with the optical cross talk probability, which is non-desirable feature for most of applications, becoming sufficiently low. The price per area has also decreased largely. As a consequence, SiPM started to be used in other IACTs including some types of cameras of SSTs and MSTs. We are considering to adopt SiPMs for a possible upgrade of the LST camera. 

In the case of LSTs, as the target energies of gamma rays are much lower than other telescopes, the contamination of the night sky background (NSB) photons to the air shower image is not negligible. Therefore, to evaluate new photo sensors, the effect of night sky background (NSB) photons must be carefully examined besides the Cherenkov photon detection efficiency.
In this context, we studied the characteristics of SiPMs from Hamamatsu photonics and we are developing a 14 pixel SiPM module that can be used for LSTs. 

 \section{Hamamatsu S13360-3075CN-UVE-1}
Among different types of SiPM we tested, here we would like to report mainly on Hamamatsu S13360-3075CN-UVE-1. 

\subsection{Sensor description}
 This sensor was developed by Hamamatsu photonics aiming for usage in IACTs. Photodetection efficiency (PDE) is tuned for near-UV wavelengths, where atmospheric Cherenkov light spectrum peaks. The most pronounced characteristic of this is its pulse width,  which will be discussed in the next sub-section. The size of the sensitive area is 3 mm x 3 mm, which consists of 1600 Geiger mode avalanche photo diode (g-APD) with a pitch of 75 um.

\subsection{PulseShape}
\label{sect:pulseshape}
\begin{figure}[th]
    \centering
\includegraphics[width=0.95\textwidth]{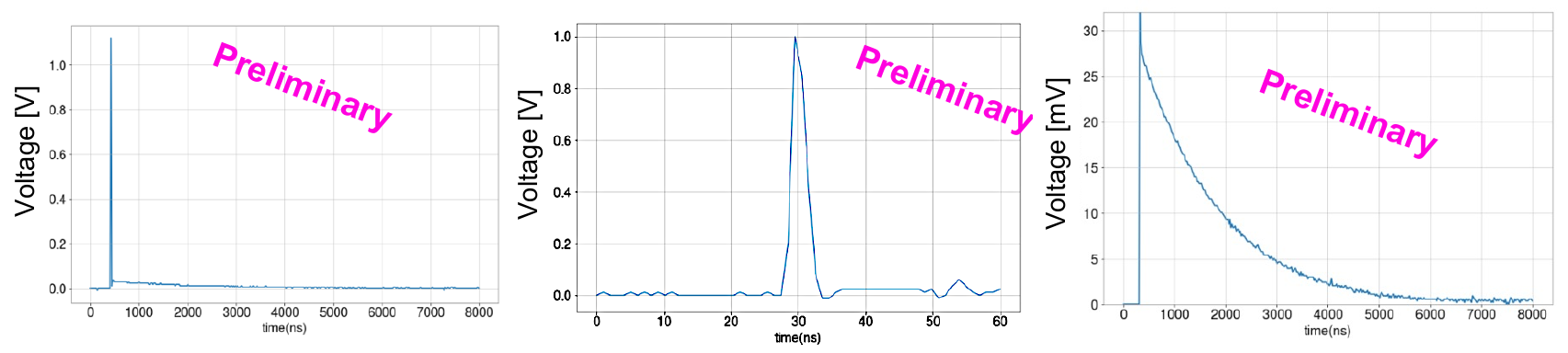}
   \caption{{\it Left}: The overall picture of the pulse shape of Hamamatsu S13360-3075CN-UVE-1. {\it Middle}: the pulse shape zoomed in the horizontal axis for the fast component. The full width half maximum is about 4 ns. {\it Right}, the pulse shape zoomed in the vertical axis for the slow component. It has an exponential shape and the decay constant is about 2 $\mu$s.}
   \label{fig:pulseshape}
\end{figure}

The pulse shape of this SiPM is shown in figure \ref{fig:pulseshape}. There are a very fast component and a very slow component as you can see in the middle and the left panel of figure \ref{fig:pulseshape}. The full width half maximum of the fast component is about 4 ns, while the slow component exhibits exponential decay with time constant of about 2 $\mu$s.

 Given that the typical duration of atmospheric Cherenkov flash from air showers is a few to several ns, 
one can extract the charge only  from the fast component to record the shower image.
However, it would be important to understand the performance of the sensor if the next photon hit the g-APD during the long tail. In section \ref{sect:recovery}, we will study the recovery time.

\subsection{Basic Performance}
\begin{figure}
    \centering
\includegraphics[width=0.70\textwidth, angle=90]{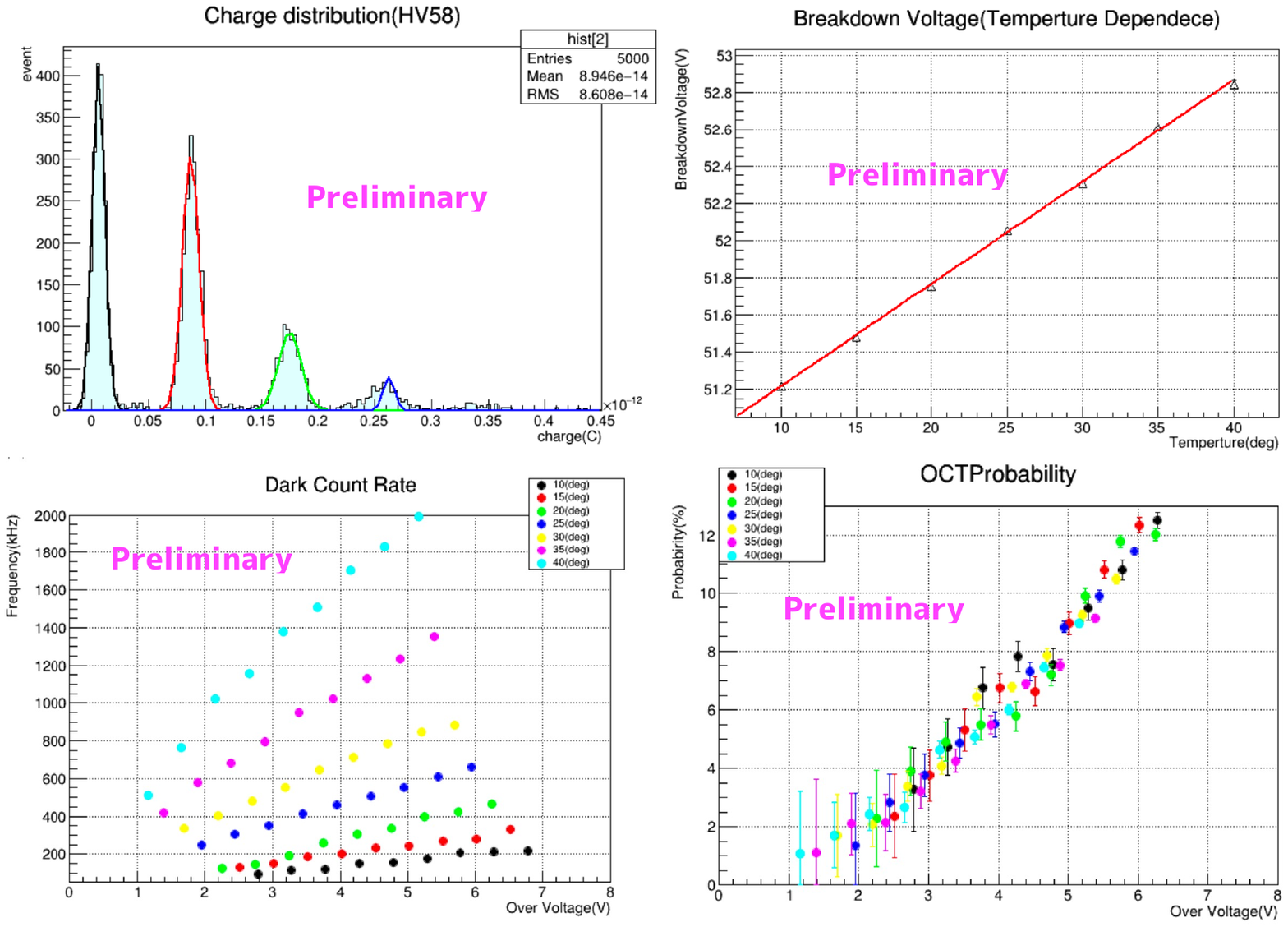}
   \caption{Basic performance of Hamamatsu S13360-3075CN-UVE-1. {\it Top Left}: Output charge disbribution for  weak input signals. Clearly separated distributions up to 4 p.e. can be seen. {\it Top right}: Temperature dependence of the break down voltage. It is about 0.053 V/$^\circ$C. {\it Bottom left}: Temperature and over-voltage dependence of the dark count rate. {\it Bottom right}: Temperature and over-voltage dependence of the optical cross talk probability. }\label{fig:Basic}
\end{figure}

    The basic performance such as charge resolution, the temperature dependence of break down voltage, the dark count rate (DCR) as a function of over voltage and temperature and the optical crosstalk probability as a function of the over voltage and temperature were studied and results are shown in figure \ref{fig:Basic}. Temperature dependence of break down voltage is about 0.053 V/$^\circ$C. This dependency may require a simple temperature compensation circuit when it is used in IACT. DCR is highly temperature and over-voltage dependent, and it is about 400 kHz at 25$^\circ$C with the over voltage of 3 volts. It is negligible compared with NSB for LSTs (see section \ref{sect:NSB}).  Optical cross talk probably also depends on the over voltage and about 4\% at 3~V, which is also acceptable for IACT application. 

\subsection{Recovery Time}
\label{sect:recovery}
\begin{figure}[h]
    \centering
\includegraphics[width=0.45\textwidth]{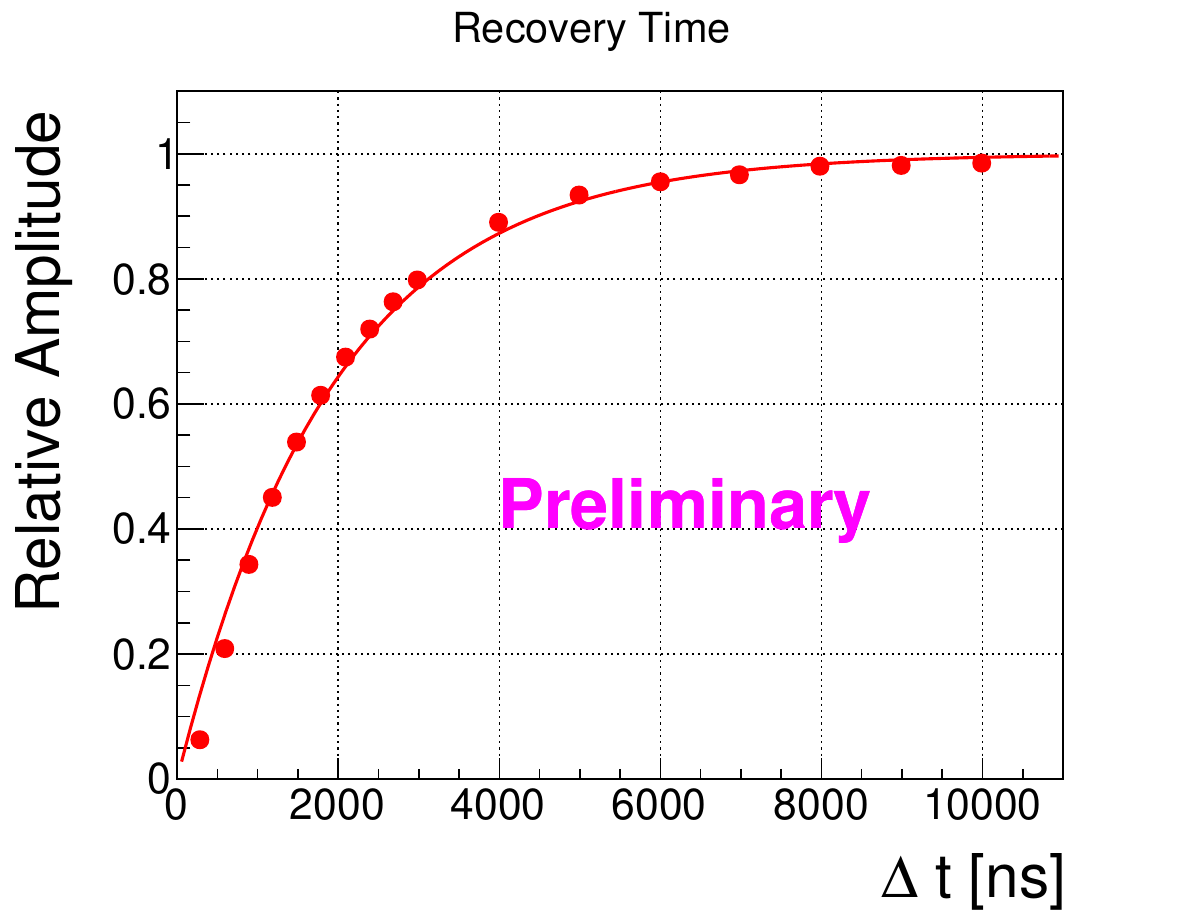}
   \caption{Relative pulse amplitude as a function of the time since the previous avalanche process. The curve can be fitted with $1 - \rm{exp}(t/\tau)$ and $\tau$ is approximately 2 \textmu s}\label{fig:recovery}
\end{figure}

Given that this sensor has a 2 \textmu s long tail in the pulse shape as shown in section \ref{sect:pulseshape}, we studied the recovery time
of this sensor. The method is the following. First we illuminate the sensor with a intense laser pulse such that all 1600 g-APDs fire. Soon after that, the second laser flashes it again and record the signal pulse amplitude.
If the time difference between the two flashes ($\Delta t$) is small, the amplitude is reduced because g-APDs are not sufficiently charged up.
The relation between the amplitude and $\Delta t$ is shown in the figure \ref{fig:recovery}.
The relation can be expressed in $1 - \rm{exp}(t/\tau)$ and $\tau$ is approximately 2 \textmu s, which is similar to the time constant of the slow component in the pulse shape.

\subsection{Effect of NSB}
 \label{sect:NSB}
\begin{figure}[h]
\centering
\includegraphics[width=0.45\textwidth]{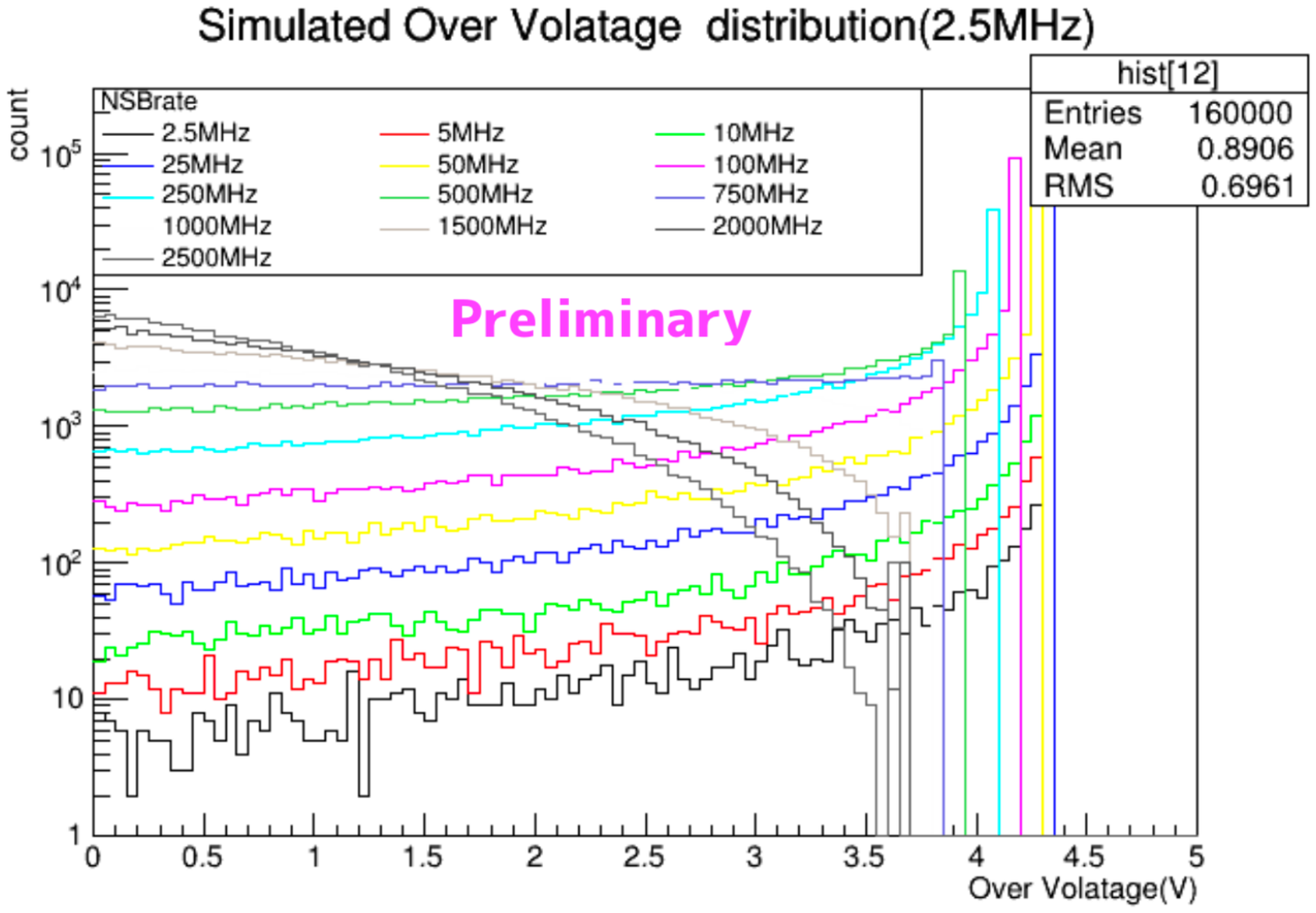}
\includegraphics[width=0.45\textwidth]{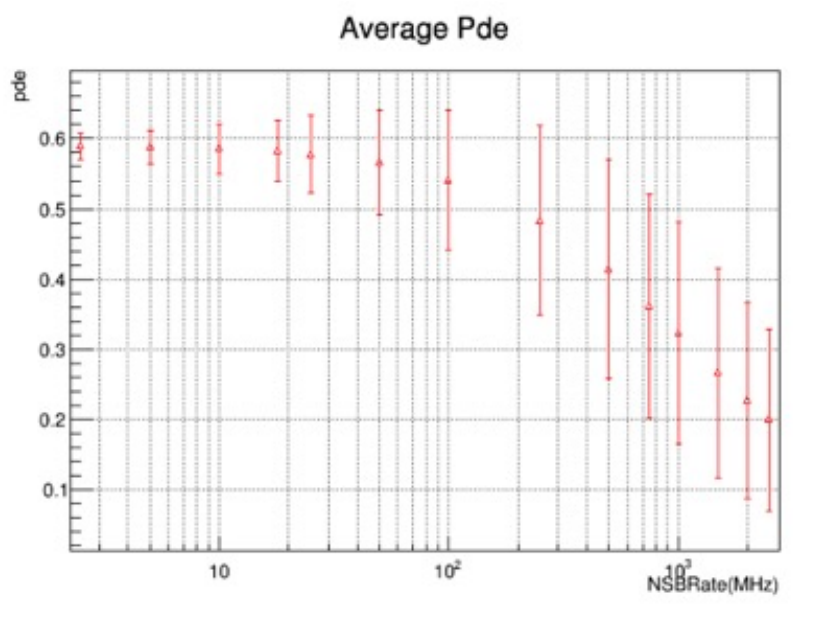}
   \caption{{\it Left}: Simulated distribution of over voltage over the 1600 g-APDs under different NSB rates. {\it Right}: The mean and the standard deviation the PDE distribution as a function the NSB rates. Three dashed vertical lines indicate the NSB rate of the dark night, 10 time the dark night and 100 times the dark night. }
   \label{fig:OVdistribution}
\end{figure}

The recovery time $\sim 2$ \textmu s may not be a problem if NSB photon detection rate is sufficiently low.
However, if the NSB level is high, significant fraction of g-APDs have a reduced over voltage at any moment.
The left panel of figure \ref{fig:OVdistribution} shows the simulated over-voltage distribution over the 1600 g-APDs at a given moment under different NSB conditions, based on numerical simulations. As NSB rate goes higher, large fraction of g-APDs have lower voltages.

The right panel of the figure \ref{fig:OVdistribution} 
 shows the drop of average PDE  as a function of the NSB level, based on numerical simulations.
As the lower over-voltage leads to lower PDE, the rate of NSB directly affects the PDE.
The rate of NSB in per 3 mm $\times$ 3 mm area of this sensor in a dark night is expected to be about 15 MHz.
The effect is negligible for dark nights, but if the rate is 10 times larger the dark night, about 10\% degradation is expected.
If it is 100 times larger, which is expected for the Moon night observations, the PDE will be less than half of the dark condition.

\begin{figure}[h]
\centering
\includegraphics[width=0.45\textwidth]{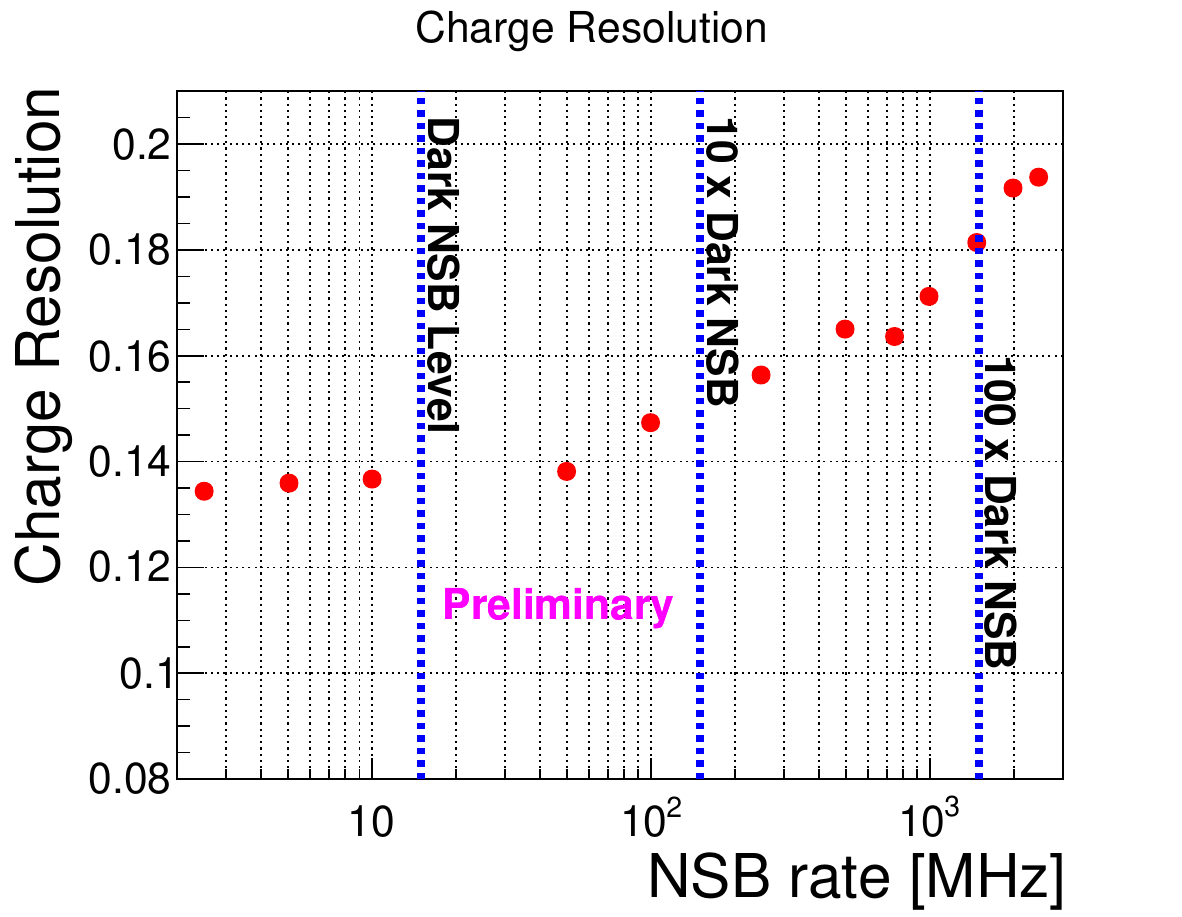}
\includegraphics[width=0.45\textwidth]{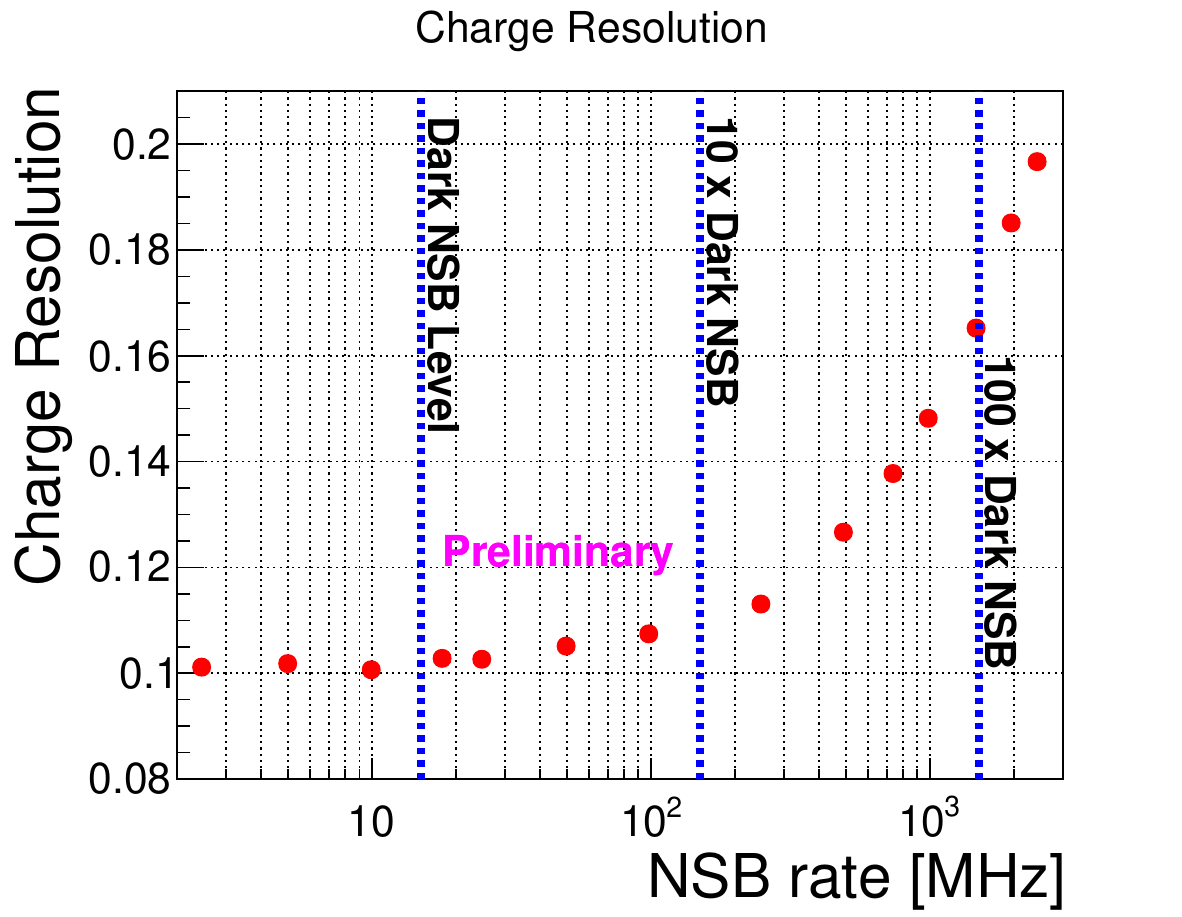}
   \caption{Charge resolution as a function of NSB rate. Measurements (left) and numerical simulation (right). Three dashed vertical lines in the right panel indicate the NSB rate of the dark night, 10 time the dark night and 100 times the dark night.}
   \label{fig:ResDrop}
\end{figure}

 Also, the variance of over voltages among g-APDs indicates the degradation of charge resolution. It is expected by simulation and confirmed by the measurements as shown in the figure \ref{fig:ResDrop}. The resolution worsen by 50\% under NSB 100 times stronger than the dark nights.

\subsection{Conclusion for Hamamatsu S13360-3075CN-UVE-1}
The $\sim 4$ ns pulse width is very attractive for LSTs, as one can suppress the contamination of NSB photons in the air shower images. However, this fast pulse is accompanied with a very long tail and a slow recovery, which results in the effective degradation of PDE and charge resolution under strong NSB. One of the advantages of SiPM with respect to the conventional PMTs are robustness of the sensor under strong illumination. For example, SiPM makes it possible to operate IACTs under the strong Moon. But the loss of PDE and charge resolution under strong background rates compromises this advantage.
The decision to adopt this sensor depends on the scientific objectives of LSTs and their priorities.
Further studies and discussions are needed.

\section{Light Guide}
\begin{figure}[h]
\centering
\includegraphics[width=0.3\textwidth]{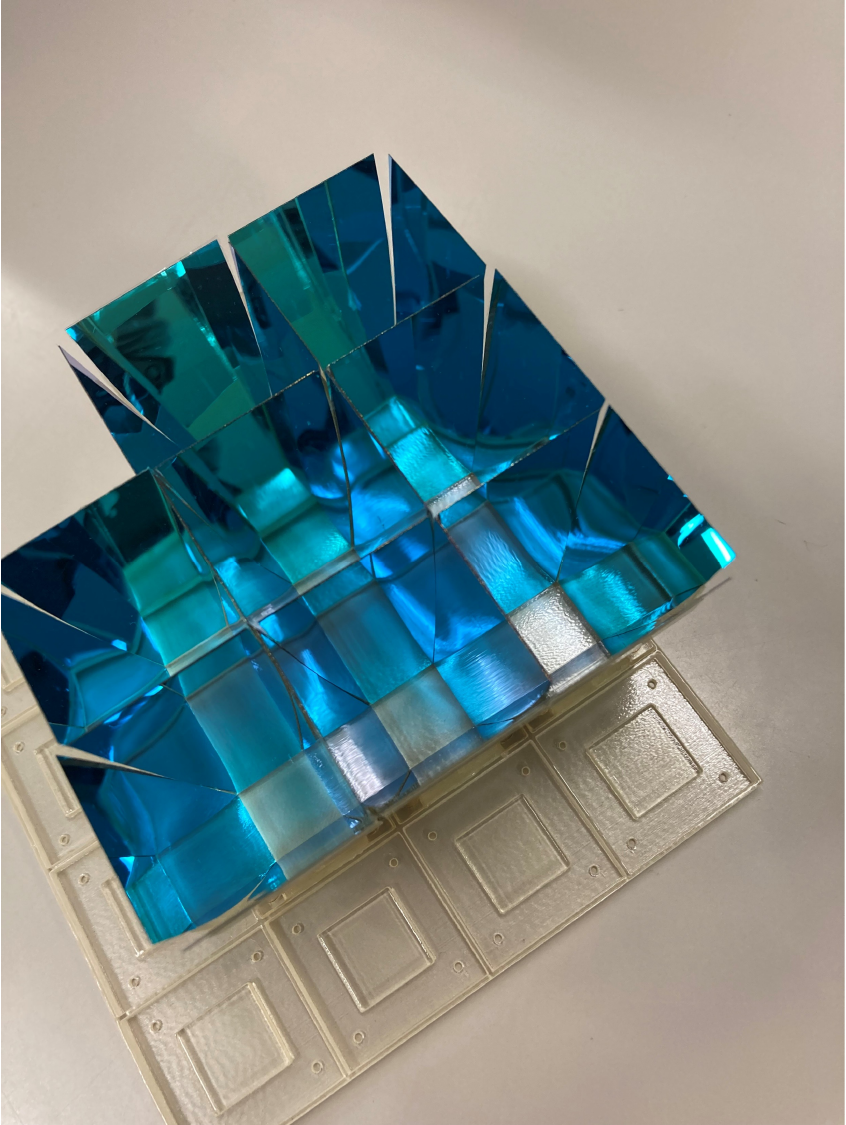}
\includegraphics[width=0.62\textwidth]{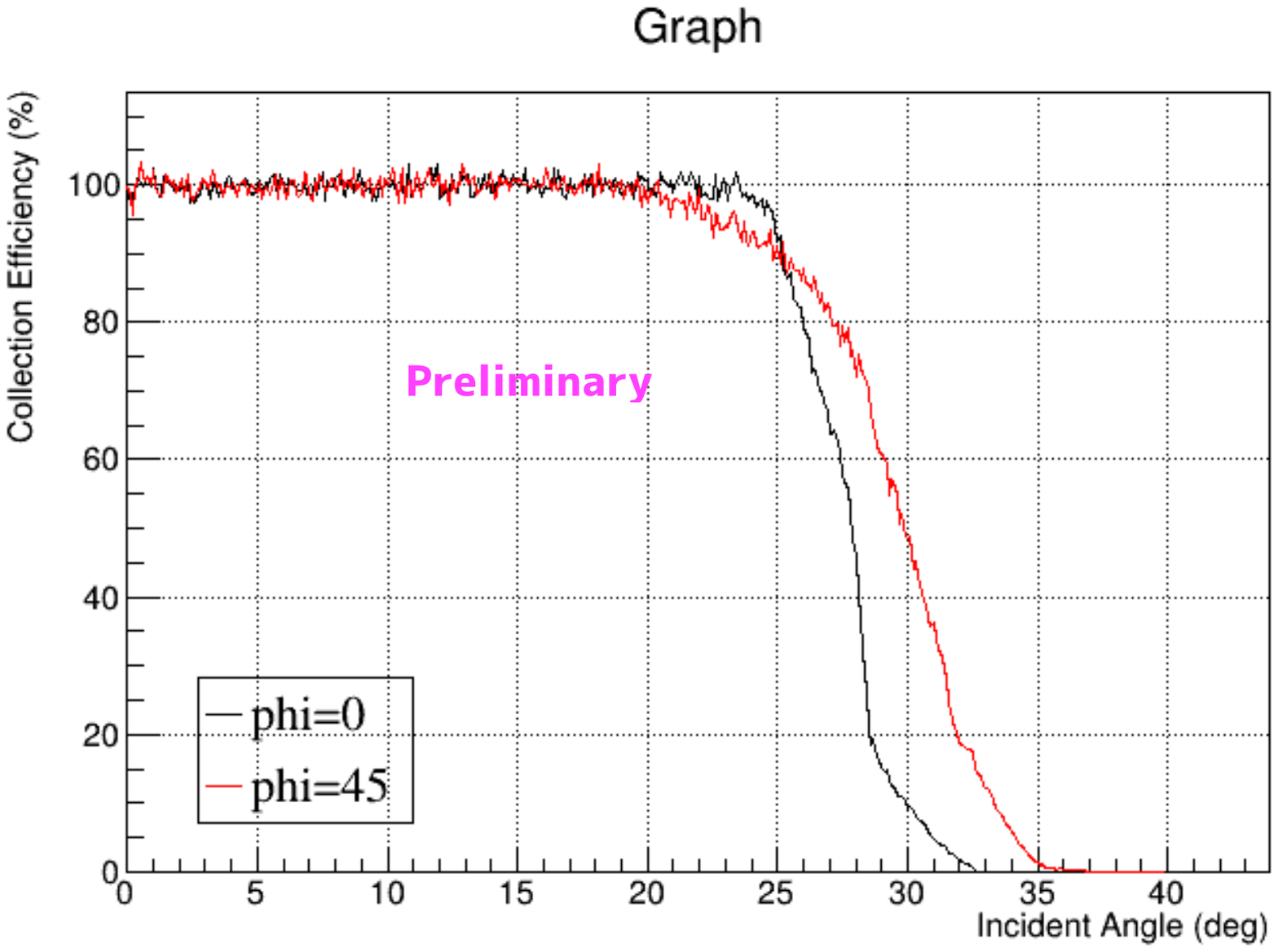}
   \caption{{\it Left}: The photo of LGs for the LST SiPM modules. Due to the coating, the reflective surface look blue. {\it Right}: Simulated collection efficiency as a function of the incident angle. Azimuthal angle 0 degree (parallel to one of the edges) and 45 degrees are shown. 
   }\label{fig:LG}
\end{figure}

As can be seen in the previous section, the reduction of NSB detection efficiency is very important if we use SiPMs in LSTs. For this purpose, a special coating on the light guides are studied. The details are presented here \cite{Okumura}. As you can see in the photo in figure \ref{fig:LG}, it reflects bluish light well while red light is suppressed. 
 The entrance and exit have a square shape. For such a case, the parabola curve used in the so-called Winstone Cone \cite{WC} is not the best option. The shape was optimized by using cubic Bezier curves following the method described in \cite{OKUMURA201218}. Collection efficiency as a function of incident angles estimated with the ray tracing simulation is shown in the right panel of figure \ref{fig:LG}. The efficiency is perfect until $20 -25$ degrees (depending on the azimuthal angles) and then sharply falls. Only the photons from the mirror dish would be collected onto the SiPM surface.

 \section{Design of the SiPM module}
\begin{figure}[h]
    \centering
\includegraphics[width=0.9\textwidth]{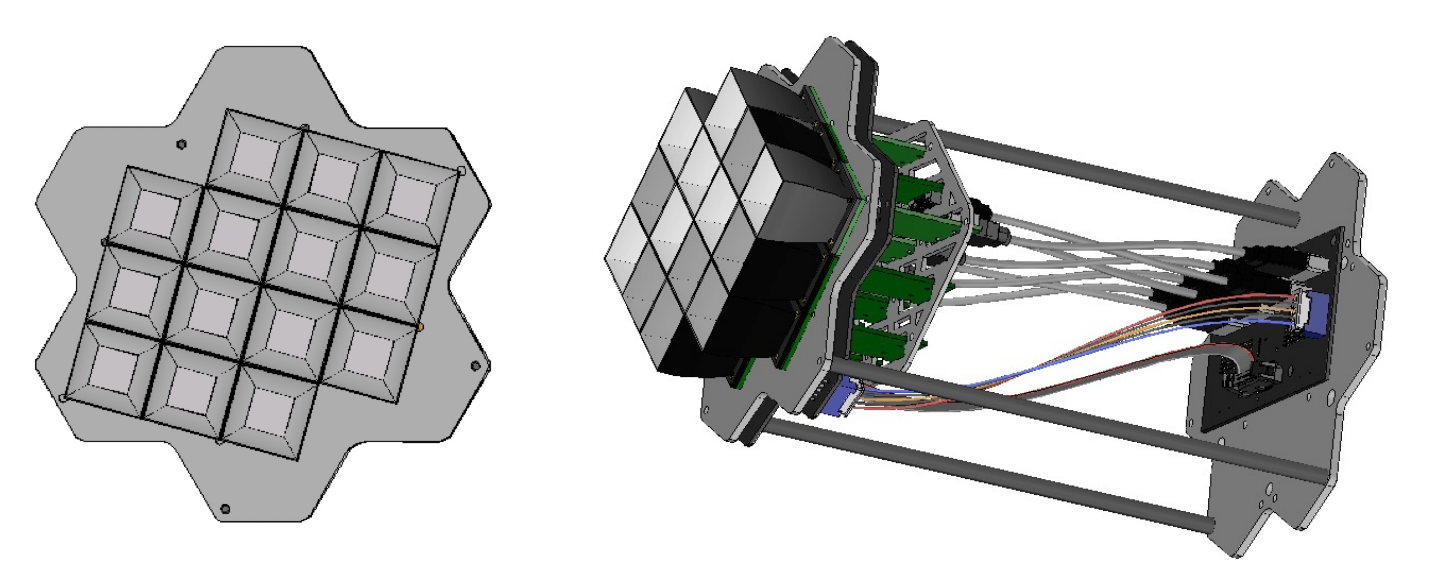}
   \caption{Left: The front view of the modules. 15 SiPM are aligned such that they do not interfere with existing PMT pixels. Right: A 3D rendering of the front part of the SiPM test bench module. The readout board will connected to the right hand side of this picture.  }
   \label{fig:module}
\end{figure}

Applications of SiPM to a large size IACT are not well studied so far. Especially, the effect of high and changing NSB rate together with temperature dependence of the sensor performance could encompass hidden problems in SiPM operation. To test them, we are developing a SiPM test bench module that can be installed in the present LST camera. Figure \ref{fig:module} shows the completion rendering of the module. It has 14 SiPM pixels with square shapes. The they are aligned such that the module can be inserted without interfering with existing PMT modules. Basically the same readout board can be used as the current 7 PMT modules. The only change needed was a change in the electronics to replace the 7 low-gain PMT readout channels with a second set of 7 high-gain channels.
 
\section{Conclusion}
Hamamatsu S13360-3075CN-UVE-1, which exhibits a fast pulse, is an attractive option for LST cameras, but the slow recovery time needs to be carefully examined since the performance degrades under strong NSB. The development of test bench modules including square shaped light guides, that can be installed in the existing LST camera, is progressing well.

\bibliographystyle{unsrt}
\bibliography{skeleton}

\begin{center}

{\bf LST Acknowledgements }

\bigskip\bigskip
{\small
We gratefully acknowledge financial support from the following agencies and organisations:

\bigskip
Conselho Nacional de Desenvolvimento Cient\'{\i}fico e Tecnol\'{o}gico (CNPq), Funda\c{c}\~{a}o de Amparo \`{a} Pesquisa do Estado do Rio de Janeiro (FAPERJ), Funda\c{c}\~{a}o de Amparo \`{a} Pesquisa do Estado de S\~{a}o Paulo (FAPESP), Funda\c{c}\~{a}o de Apoio \`{a} Ci\^encia, Tecnologia e Inova\c{c}\~{a}o do Paran\'a - Funda\c{c}\~{a}o Arauc\'aria, Ministry of Science, Technology, Innovations and Communications (MCTIC), Brasil;
Ministry of Education and Science, National RI Roadmap Project DO1-153/28.08.2018, Bulgaria;
Croatian Science Foundation, Rudjer Boskovic Institute, University of Osijek, University of Rijeka, University of Split, Faculty of Electrical Engineering, Mechanical Engineering and Naval Architecture, University of Zagreb, Faculty of Electrical Engineering and Computing, Croatia;
Ministry of Education, Youth and Sports, MEYS  LM2015046, LM2018105, LTT17006, EU/MEYS CZ.02.1.01/0.0/0.0/16\_013/0001403, CZ.02.1.01/0.0/0.0/18\_046/0016007 and CZ.02.1.01/0.0/0.0/16\_019/0000754, Czech Republic; 
CNRS-IN2P3, the French Programme d’investissements d’avenir and the Enigmass Labex, 
This work has been done thanks to the facilities offered by the Univ. Savoie Mont Blanc - CNRS/IN2P3 MUST computing center, France;
Max Planck Society, German Bundesministerium f{\"u}r Bildung und Forschung (Verbundforschung / ErUM), Deutsche Forschungsgemeinschaft (SFBs 876 and 1491), Germany;
Istituto Nazionale di Astrofisica (INAF), Istituto Nazionale di Fisica Nucleare (INFN), Italian Ministry for University and Research (MUR);
ICRR, University of Tokyo, JSPS, MEXT, Japan;
JST SPRING - JPMJSP2108;
Narodowe Centrum Nauki, grant number 2019/34/E/ST9/00224, Poland;
The Spanish groups acknowledge the Spanish Ministry of Science and Innovation and the Spanish Research State Agency (AEI) through the government budget lines PGE2021/28.06.000X.411.01, PGE2022/28.06.000X.411.01 and PGE2022/28.06.000X.711.04, and grants PID2022-139117NB-C44, PID2019-104114RB-C31,  PID2019-107847RB-C44, PID2019-104114RB-C32, PID2019-105510GB-C31, PID2019-104114RB-C33, PID2019-107847RB-C41, PID2019-107847RB-C43, PID2019-107847RB-C42, PID2019-107988GB-C22, PID2021-124581OB-I00, PID2021-125331NB-I00; the ``Centro de Excelencia Severo Ochoa" program through grants no. CEX2019-000920-S, CEX2020-001007-S, CEX2021-001131-S; the ``Unidad de Excelencia Mar\'ia de Maeztu" program through grants no. CEX2019-000918-M, CEX2020-001058-M; the ``Ram\'on y Cajal" program through grants RYC2021-032552-I, RYC2021-032991-I, RYC2020-028639-I and RYC-2017-22665; the ``Juan de la Cierva-Incorporaci\'on" program through grants no. IJC2018-037195-I, IJC2019-040315-I. They also acknowledge the ``Atracción de Talento" program of Comunidad de Madrid through grant no. 2019-T2/TIC-12900; the project ``Tecnologi\'as avanzadas para la exploracio\'n del universo y sus componentes" (PR47/21 TAU), funded by Comunidad de Madrid, by the Recovery, Transformation and Resilience Plan from the Spanish State, and by NextGenerationEU from the European Union through the Recovery and Resilience Facility; the La Caixa Banking Foundation, grant no. LCF/BQ/PI21/11830030; the ``Programa Operativo" FEDER 2014-2020, Consejer\'ia de Econom\'ia y Conocimiento de la Junta de Andaluc\'ia (Ref. 1257737), PAIDI 2020 (Ref. P18-FR-1580) and Universidad de Ja\'en; ``Programa Operativo de Crecimiento Inteligente" FEDER 2014-2020 (Ref.~ESFRI-2017-IAC-12), Ministerio de Ciencia e Innovaci\'on, 15\% co-financed by Consejer\'ia de Econom\'ia, Industria, Comercio y Conocimiento del Gobierno de Canarias; the ``CERCA" program and the grant 2021SGR00426, both funded by the Generalitat de Catalunya; and the European Union's ``Horizon 2020" GA:824064 and NextGenerationEU (PRTR-C17.I1).
State Secretariat for Education, Research and Innovation (SERI) and Swiss National Science Foundation (SNSF), Switzerland;
The research leading to these results has received funding from the European Union's Seventh Framework Programme (FP7/2007-2013) under grant agreements No~262053 and No~317446;
This project is receiving funding from the European Union's Horizon 2020 research and innovation programs under agreement No~676134;
ESCAPE - The European Science Cluster of Astronomy \& Particle Physics ESFRI Research Infrastructures has received funding from the European Union’s Horizon 2020 research and innovation programme under Grant Agreement no. 824064.
}
\end{center}



%
%
%

\end{document}